\documentclass[prl,preprint,showpacs,preprintnumbers,amsmath,amssymb,superscriptaddress]{revtex4}

\usepackage{graphicx}
\usepackage{graphics}
\usepackage{epsfig}

\begin{document}

\bibliographystyle{prsty} 

\title{Spin Injection and Relaxation in Ferromagnet-Semiconductor Heterostructures}

\author{C. Adelmann}
\affiliation{Department of Chemical Engineering and Materials
Science,}
\author{X. Lou}
\affiliation{School of Physics and Astronomy,\\University of
Minnesota, Minneapolis, MN 55455}
\author{J. Strand}
\affiliation{School of Physics and Astronomy,\\University of
Minnesota, Minneapolis, MN 55455}
\author{C. J. Palmstr{\o}m}
\affiliation{Department of Chemical Engineering and Materials
Science,}
\author{P. A. Crowell}\email{crowell@physics.umn.edu}
\affiliation{School of Physics and Astronomy,\\University of
Minnesota, Minneapolis, MN 55455}
\begin{abstract}\
We present a complete description of spin injection and detection
in Fe/Al$_x$Ga$_{1-x}$As/GaAs heterostructures for temperatures
from 2 to 295~ K.   Measurements of the steady-state spin
polarization in the semiconductor indicate three temperature
regimes for spin transport and relaxation. At temperatures below
70~K, spin-polarized electrons injected into quantum well
structures form excitons, and the spin polarization in the quantum
well depends strongly on the electrical bias conditions. At
intermediate temperatures, the spin polarization is determined
primarily by the spin relaxation rate for free electrons in the
quantum well.  This process is slow relative to the excitonic spin
relaxation rate at lower temperatures and is responsible for a
broad maximum in the spin polarization between 100 and 200~K. The
spin injection efficiency of the Fe/Al$_x$Ga$_{1-x}$As Schottky
barrier decreases at higher temperatures, although a steady-state
spin polarization of at least 6\% is observed at 295 K.

\end{abstract}


\maketitle


Ferromagnetic metals such as iron are natural sources of
spin-polarized electrons, and semiconductors have been shown to be
an ideal host for the transport and manipulation of spin.  The
demonstration of electrical spin injection from conventional
ferromagnetic
metals\cite{Zhu_2001,Hanbicki_2002,Motsnyi_2002,Jiang_2003,Strand_2003}
has addressed the possibility of purely electronic control of spin
transport in semiconductors. For example, the steady-state spin
polarization electrically injected into a quantum well from an
Fe/Al$_x$Ga$_{1-x}$As Schottky contact has been shown to be as
high as 32~\% at 2~K\cite{Hanbicki_2003}. Improved efficiencies
have been achieved for injection through an artificial tunnel
barrier\cite{vantErve_2004,Jiang_2004}, and evidence for
electrical spin injection at room temperature has been
reported\cite{Zhu_2001,Motsnyi_2003}. In spite of these successes,
no experiment on ferromagnet-semiconductor heterostructures has
addressed the properties of these devices over a wide range of
temperatures and electrical bias conditions.

In this Letter we report on a comprehensive study of spin
injection in Fe/Al$_x$Ga$_{1-x}$As/GaAs heterostructures from 2~K
to 295~K.   When a shallow GaAs quantum well (QW) is used as a
spin detector, three distinct temperature regimes for spin
transport and relaxation are identified.  Below 70~K, the bias
dependence of the spin polarization in the QW is clearly
influenced by excitonic effects. A pronounced peak appears in the
steady-state polarization over a narrow bias range.  This peak
disappears rapidly with increasing temperature. Between 70 and
150~K, the spin polarization {\it increases} with temperature over
a wide range of bias voltages.  We show that the temperature
dependence of the polarization signal from 2 to 150 K can be
understood in terms of a crossover from excitonic to free electron
spin relaxation in the quantum well. Above 180~K, the steady-state
spin polarization decreases in all heterostructures that we have
studied but is at least 6\% at 295~K.  Measurements using a bulk
GaAs spin detector indicate that the decrease at higher
temperatures is due in part to a reduction in the spin injection
efficiency of the Schottky barrier.

\begin{figure}
    \includegraphics{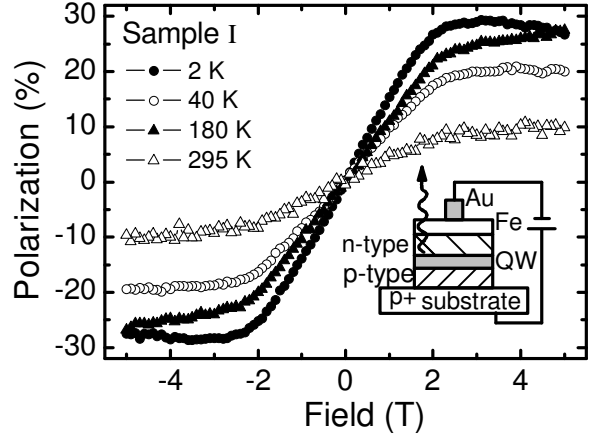}
    \caption{Electroluminescence polarization (ELP) as a function of
    magnetic field for sample I at the
    temperatures and bias voltages indicate in the legend. A schematic of
    the structure for samples I and II is shown in the inset.  The quantum well
    is omitted in sample III.}
    \label{fig:Faraday}
\end{figure}

Each of the epitaxial ferromagnet-semiconductor heterostructures
used for these measurements consists of a Schottky diode in series
with a {\it n-i-p} junction\cite{Zhu_2001,Hanbicki_2002}. The
design of the Schottky tunnel barrier follows the approach of
Hanbicki {\it et al.}\cite{Hanbicki_2003}. Three samples will be
discussed in detail in this paper.  The first two, denoted  I and
II, use quantum wells as optical detectors. Sample~I is grown on a
p-type ($p=1\times 10^{18}$~cm$^{-3}$) GaAs (100) substrate and
consists of 300~nm p-GaAs ($p=1\times 10^{17}$~cm$^{-3}$), 150~nm
p-Al$_{0.1}$Ga$_{0.9}$As ($p=1\times 10^{16}$~cm$^{-3}$), 25~nm
i-Al$_{0.1}$Ga$_{0.9}$As, 10~nm i-GaAs QW, 25~nm
i-Al$_{0.1}$Ga$_{0.9}$As, followed by a 100~nm
n-Al$_{0.1}$Ga$_{0.9}$As ($n=1\times 10^{16}$~cm$^{-3}$)~drift
layer. The Schottky junction is then formed by growing a
$n\rightarrow n^+$ transition layer going from $n = 1\times
10^{16}$~cm$^{-3}$ up to $5\times 10^{18}$~cm$^{-3}$ over a
thickness of 15~nm.  This is followed by 15~nm
n$^+$-Al$_{0.1}$Ga$_{0.9}$As ($n^+=5\times 10^{18}$~cm$^{-3}$),
5~nm Fe, and a 2.5~nm Al capping layer. The Fe and Al layers are
grown at a temperature of 0~$^\circ$C. Sample II is identical to
sample~I except for a lower doping ($p=3\times 10^{15}$~cm$^{-3}$)
in the p-Al$_{0.1}$Ga$_{0.9}$As layer immediately beneath the QW
structure. Sample~III differs from sample~I only in that the 10~nm
i-GaAs QW is eliminated.  The optical emission from this sample is
dominated by GaAs band-edge luminescence emitted from the
substrate. The samples are processed into light-emitting-diodes by
photolithography and wet etching. After processing, each device is
annealed at 250~$^\circ$C in a N$_2$ atmosphere for one hour.  A
schematic of a sample is shown in the inset of
Fig.~\ref{fig:Faraday}.  Light is collected through the top of the
device.

The spin detection measurements are carried out using the
electroluminescence polarization (ELP)~technique in the Faraday
geometry\cite{ELP}.  Light is emitted by electrons that tunnel
into the semiconductor from the Fe film and recombine with
unpolarized holes from the substrate.  The electroluminescence
polarization $P=(I_+ - I_-)/(I_+ + I_-)$, where $I_+$ and $I_-$
are the intensities of right and left circularly polarized light,
is measured as a function of magnetic field, temperature, and the
bias voltage across the device. For samples I and II, the
electroluminescence (EL) at low temperatures is dominated by the
QW heavy-hole exciton, for which $P$ is equal to the steady-state
electron spin polarization in the QW. The polarization for these
samples below 200~K is determined from the intensities integrated
over a window 3~meV wide surrounding the heavy-hole exciton peak.
At higher temperatures, the electroluminescence from samples I and
II becomes dominated by recombination in the substrate and the
data are windowed over a 5~meV window around the EL maximum.  The
EL from sample III is due to band-edge recombination in GaAs at
all temperatures, and in this case $P$, which is determined from
the spectrum integrated over a 40~meV window, is expected to be
equal to half of the steady-state spin polarization in the
detection region\cite{OO}. Only the raw optical polarization $P$
will be shown in this paper. This includes small contributions
from magneto-absorption in the Fe film (less than 2\% in all cases
discussed here) and, at very low temperatures, the Zeeman
splitting of electron and hole states in the semiconductor.

The electroluminescence polarization $P$ for sample I is shown as
a function of magnetic field in Fig.~\ref{fig:Faraday}.  The data
are obtained at temperatures ranging from 2~K to 295~K at the bias
voltages indicated in the legend. As demonstrated in previous
work\cite{Zhu_2001,Hanbicki_2002}, $P$ is approximately
proportional to the magnetization of iron, which saturates at an
applied field of $H= 4\pi M = 2.1$~T.  This magnetic field
dependence is observed for all three samples.  For samples I and
II, a polarization of 8\% at 2.5~T (6\% after background
subtraction)~is observed even at 295~K.

\begin{figure}
    \includegraphics{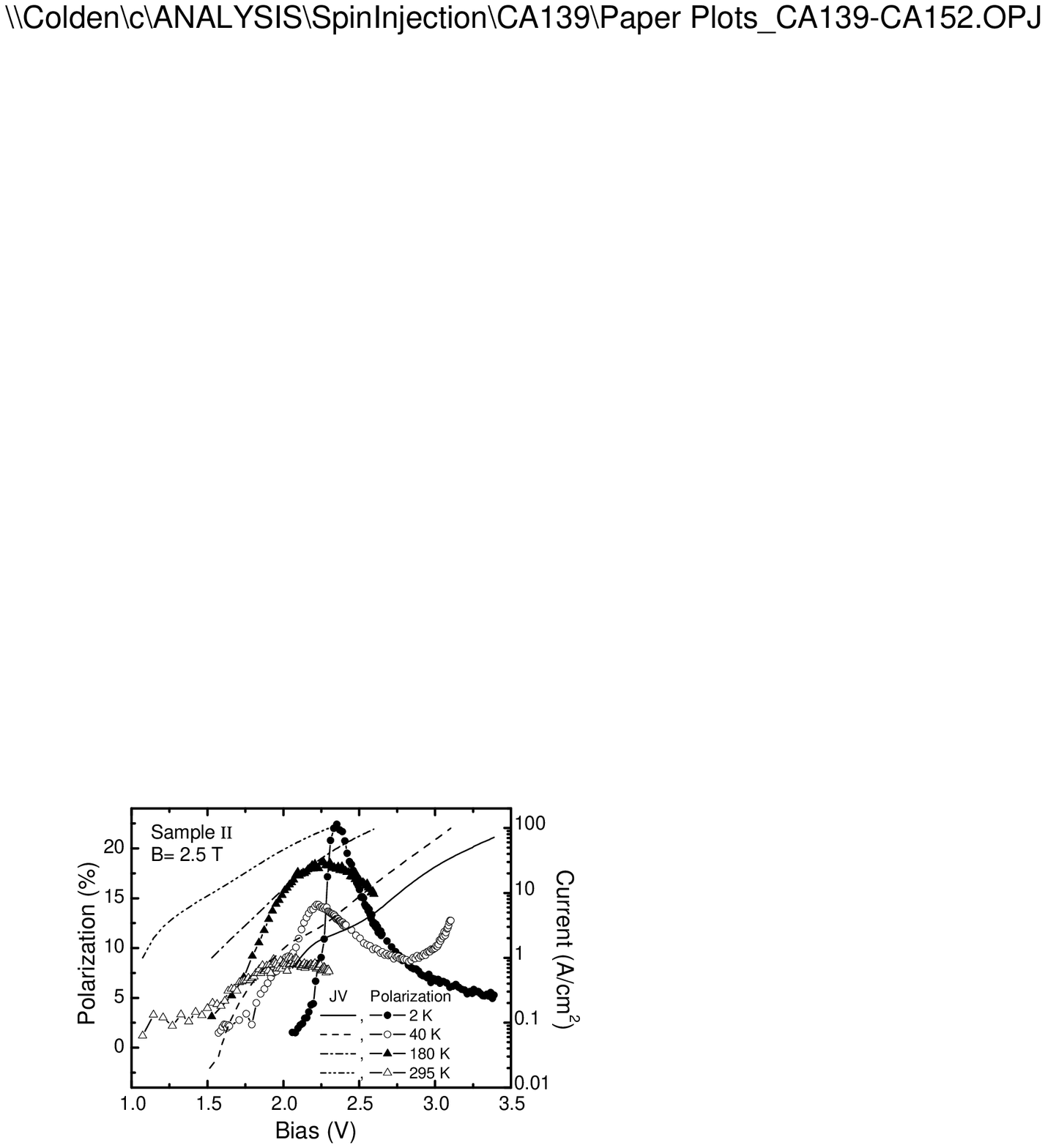}
    \caption{The polarization (symbols) is shown as a function of the bias voltage for
    sample II in a field of 2.5~T at the temperatures
    indicated in the legend.  The curves are the
    corresponding current-voltage characteristics.}
    \label{fig:Bias1}
\end{figure}

A complete picture of the spin transport properties of these
devices can be obtained by measuring the optical polarization as a
function of the bias voltage between the ferromagnetic electrode
and the substrate.  For this measurement, the magnetic field is
held fixed at 2.5~T, just above the saturation field of Fe.
Results at several different temperatures are shown for sample II
in Fig.~\ref{fig:Bias1}.
These data show three distinguishing features.  The first is the
pronounced peak in the polarization as a function of bias that is
observed at 2~K. Second, the maximum polarization at 180~K is {\it
higher} than that measured at 40~K. Finally, there is a
significant decrease in the polarization signal between 180 and
295~K.

\begin{figure}
    \includegraphics{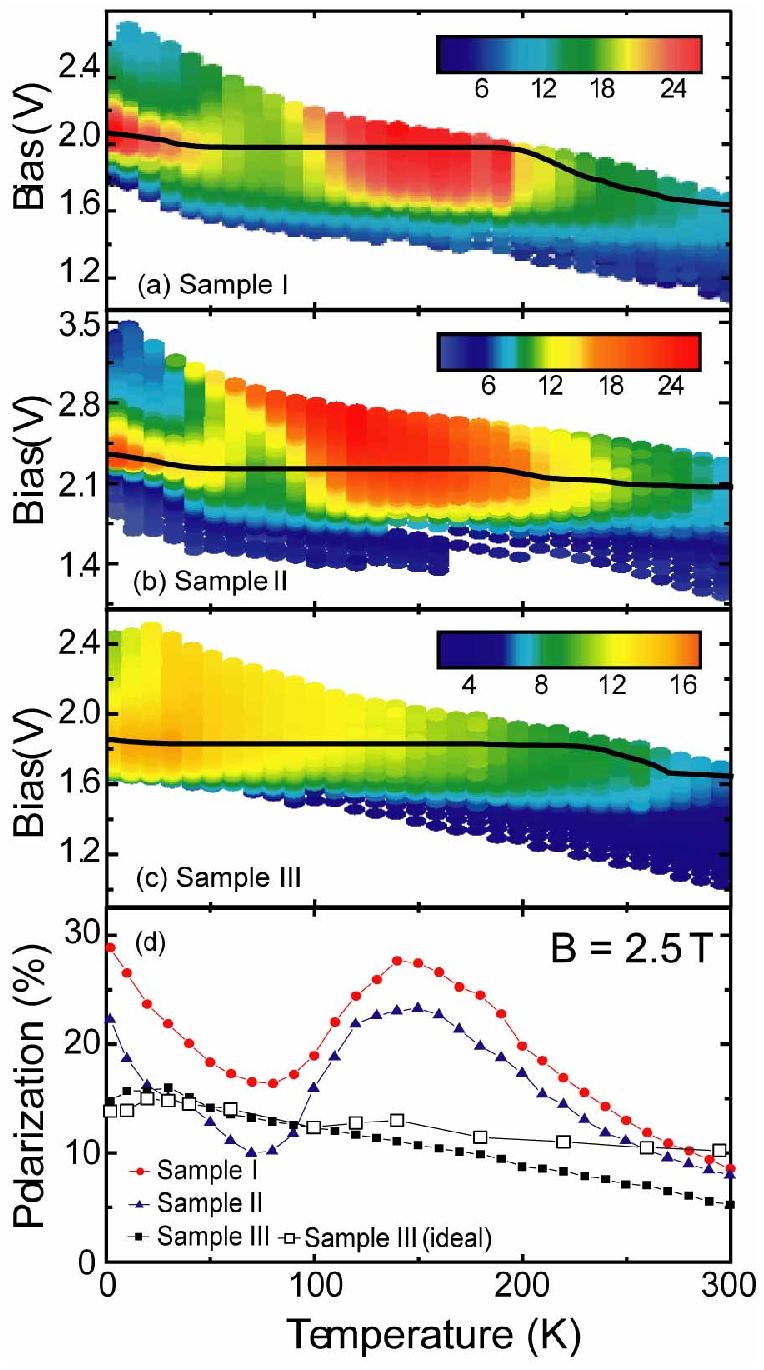}
    \caption{(color) (a),(b),(c) The polarization measured at the
    electroluminescence peak
    is shown as a function of the temperature and bias voltage for
    samples I, II, and III in a field of 2.5~T.  The color scales
    are indicated in each panel.
   (d) Optical polarization (closed symbols) is shown as a
   function of temperature for each of the samples in this study. The data
   are shown at points along the solid black curves in panels (a)-(c).
   The maximum polarization expected for sample III for the ideal
   case of 100\% injection efficiency
   is shown using open symbols. This is based on the calibration
   procedure described in the text.}
    \label{fig:Bias2}
\end{figure}

It is evident from Fig.~\ref{fig:Bias1} that the temperature and
bias dependence of the polarization signal are complex. Complete
maps of the polarization as a function of temperature and bias
voltage are provided in Figs.~\ref{fig:Bias2}(a-c) for the three
samples discussed in this paper.  The closed symbols in
Fig.~\ref{fig:Bias2}(d) show the polarization
at the voltages along the solid curves in each of the first three
panels.  The data in Fig.~\ref{fig:Bias2}(d) approximate the
maximum polarization at each temperature.

It is clear from Fig.~\ref{fig:Bias2} that there are two regions
of maximum polarization signal for QW detectors. The first is at
low temperature over a narrow bias range.  The second maximum
occupies a much wider bias range at intermediate temperatures,
between 70 and 200~K.  For the bulk GaAs detector of sample III,
there is a single maximum at low temperature, and the polarization
signal decreases with increasing temperature above 20~K for all
biases. The temperature dependence of the maximum polarization
that we observe for QW detectors agrees qualitatively with recent
results obtained below 100 K with an artificial tunnel barrier as
the injector\cite {Jiang_2004}.

The polarization signal $P$ measured in these experiments can be
related to the injected spin by $P = \alpha
S_i/(1+\tau_r/\tau_s)$, where $S_i$ (maximum value = $1/2$) is the
spin that is injected into the quantum well, $\tau_r$ is the
recombination time, $\tau_s$ is the spin relaxation time, and
$\alpha$ is a factor determined by the optical selection rules.
For the two QW samples below 200~K, the EL is dominated by the
heavy-hole exciton, and $\alpha = 2$.  For sample III, there is no
confinement and $\alpha =1$  at all temperatures\cite{OO}.  We
focus first on the QW samples. The fact that the polarization
signal always increases with bias near threshold can be related to
a decrease in $\tau_r$ with increasing bias, as would be expected
due to the flattening of the bands in the {\it n-i-p} junction.
The sharp peak in the response at low temperature occurs at the
bias where the ratio $\tau_r/\tau_s$ is smallest. This peak
disappears with increasing temperature because $\tau_r$ increases,
as is expected for heavy-hole excitons in shallow quantum
wells\cite{QW_recombination_time} and verified for our QW's using
Hanle effect measurements\cite{OO}.

There are, however, important features of the QW data in
Fig.~\ref{fig:Bias2} that cannot be due simply to variations in
the recombination time.  As can be seen in
Figs.~\ref{fig:Bias2}(a) and (b), the rapid decrease in $P$ from
10 - 70~K occurs only over a narrow bias range.  For higher bias
voltages, the polarization signal actually increases with
temperature from 10~K up to 150~K. These unusual effects are due
to the dependence of the spin relaxation time $\tau_s$ on bias
voltage and temperature.

The behavior between 70 and 150~K can be understood in terms of
the  D'yakonov-Perel (DP) mechanism\cite{Pikus_Titkov,DK_1986}, in
which the electron spins precess incoherently about the spin-orbit
field.  In a manner similar to motional narrowing, this process
can be suppressed if the momentum scattering time $\tau_p$ is
short enough. For the case of electron spin-relaxation in quantum
wells, $\tau_s^{-1} \propto \tau_p T$\cite{DK_1986,highT}, and so
we expect $\tau_s$ to {\it increase} with increasing temperature
if the momentum scattering time decreases with temperature faster
than $1/T$.  As noted by Jiang {\it et al.}\cite{Jiang_2004}, the
rapid onset of optical phonon scattering above 70~K therefore
provides a reasonable explanation for the increase in the
polarization signal at higher temperatures. We find that $P$ (and
hence $\tau_s$) continues to increase up to 150~K\cite{highT}.

The DP mechanism alone, however, cannot explain the temperature
and bias dependence that is observed below 70~K.  We have
considered various models that treat consistently the dependence
of the DP relaxation rate on temperature and the kinetic energy of
the injected carriers.  Most importantly, none of the common
models for free electron spin relaxation predicts the increase in
$P$ with temperature that is observed at high biases.  As noted
above, this trend starts at progressively lower temperatures (far
below the onset of optical phonon scattering) at the highest bias
voltages.  Clearly some other process besides the DP mechanism is
contributing to the electron spin relaxation at low temperatures.

The key to understanding the low-temperature behavior observed in
Figures \ref{fig:Bias1} and \ref{fig:Bias2} is the formation of
excitons.   The electron-hole exchange interaction has been shown
to enhance the spin relaxation rate significantly compared to that
observed for free
electrons\cite{Maialle_1993,Vinattieri_1994,Blackwood_1994,Munoz_1995,Gerlovin_2004}.
The exchange interaction can be tuned by controlling the spatial
overlap of the electron and hole wave functions.   For example, a
factor of five decrease in the spin relaxation rate in a GaAs QW
at 20~K was observed by Vinatierri {\it et al.}  as the electric
field was increased from 0 to 30~kV/cm\cite{Vinattieri_1994}.  Any
other parameter that decreases the electron-hole overlap, such as
an increase in temperature or a decrease in the confining
potential, should also suppress the excitonic contribution to the
electron spin relaxation rate.

The experimental situation is complicated  by the fact that the
polarization signal depends on both the recombination and spin
relaxation rates.  For this reason, it is extremely difficult to
model the full bias dependence at low temperatures. As noted
above, the sharp decrease in the {\it maximum} signal with
increasing $T$ between 2 and 70~K is consistent with the observed
increase in the excitonic recombination time.  However, the fact
that the polarization signal increases with $T$ at higher biases
is due to a crossover from excitonic spin relaxation at low
temperatures to slower free electron spin relaxation at higher
temperatures. The electron-hole overlap can be suppressed either
by increasing temperature or by increasing the electric field at
the quantum well.  An example of the latter effect can be seen in
the data for Sample II at 40~K in Fig.~\ref{fig:Bias1}.   $P$ is
actually increasing at the highest biases, for which the measured
Stark shift indicates an electric field in the QW of order
$10^4$~V/cm. Although the details of the low-temperature behavior
will depend on both $\tau_s$ and $\tau_r$, the clear boundary
separating the low-temperature regime from the broad maximum
observed at intermediate temperatures in Figs.~\ref{fig:Bias2}(a)
and (b) is associated with the suppression of the electron-hole
exchange.

We therefore find that the observed polarization signal in the
quantum well systems below 150 K can be understood in terms of a
crossover from an excitonic regime at low temperatures to the
regime above 70~K in which free electron spin relaxation by the DP
mechanism applies.  Above 150~K, however, the polarization signal
begins to decrease at all biases. This can be attributed in part
to a crossover from QW to bulk-dominated emission, but a more
fundamental question is whether the spin injection efficiency,
which we have assumed to be constant for the purposes of the
preceding discussion, decreases with increasing temperature.
Sample III, which does not have a QW, provides an opportunity to
test this assumption. In this case, recombination occurs in the
p-GaAs layer at all temperatures, and excitonic effects are
relatively weak. The maximum ELP at low temperatures is
approximately 15\%, which corresponds to a steady-state spin
polarization of 30\%.

The advantage of using a bulk recombination region is that the
ratio $\tau_r/\tau_s$ can be measured over the entire temperature
range by means of the Hanle effect\cite{OO}, thus allowing us to
calibrate the spin detector\cite{Motsnyi_2003}. From the Hanle
curve at each temperature we calculate the ideal value
$P=S_i/(1+\tau_r/\tau_s)$ of the optical polarization signal for
the case $S_i = 0.21$, which corresponds to a spin injection
efficiency of 100\% from Fe. The results are shown as the open
symbols in Fig.~\ref{fig:Bias2}(d). The relative agreement with
the results found for sample III at low temperatures suggests that
the maximum spin injection efficiency achieved with the Schottky
barrier is nearly unity.  At temperatures above 100~K the measured
values start to drop faster than the ideal case, falling 50\%
below the limiting value at room temperature.  This suggests that
some other mechanism, such as thermionic emission, contributes
significantly to the injection current above 100~K.

Our results demonstrate that the spin injection efficiency of the
Fe/Al$_x$Ga$_{1-x}$As Schottky barrier remains extremely high up
to 150~K and is of order 50\% at room temperature. The bias and
temperature dependence of the steady-state polarization  are
attributed primarily to changes in the sensitivity of the spin
detector, and steady-state spin polarizations greater than 20\%
can be reached over a large range of temperature and bias voltage.
Our discussion has ignored the possibility that the injection
efficiency itself may depend on the bias conditions, as discussed
in several theoretical
proposals\cite{Albrecht_2002,Yu_2002,Osipov_2004}. These
approaches might explain some of the extremely strong bias
dependence observed at low temperatures, but they cannot be
addressed satisfactorily until a spin detector is developed that
can be calibrated over a wide range of bias conditions.  The
experiment discussed here has identified several of the factors
that must be considered in order to achieve this goal.

This work was supported by the DARPA SPINS program, ONR, and the
University of Minnesota MRSEC (NSF DMR-0212032).  We thank J. ~Xie
for assistance with processing.


\end{document}